\documentclass[fleqn,usenatbib]{mnras}

\usepackage{graphicx,times}             
\usepackage{natbib}
\usepackage{footnote}
\usepackage[T1]{fontenc}

\DeclareRobustCommand{\VAN}[3]{#2}
\let\VANthebibliography\thebibliography
\def\thebibliography{\DeclareRobustCommand{\VAN}[3]{##3}\VANthebibliography}

\usepackage{color}
\usepackage{graphicx,amsmath}
\usepackage{rotating}
\usepackage{txfonts}
\usepackage{pdflscape,lscape}
\usepackage[T1]{fontenc}
\usepackage{aecompl}

\usepackage{graphicx,times}             
\usepackage{natbib,amsmath}

\usepackage{graphicx}	
\usepackage{amsmath}	

\newcommand{\degree}{$^{\circ}$} 
\newcommand{\km}{${\rm km}\,{\rm s}^{-1}$}

\newcommand{\hi}{H\,{\sc i}}

\newcommand{\msolar}{{\rm M}_{\odot}}
\newcommand{\re}{$R_{\rm e}$}

\title[Southern UDGs and LSBGs]{HIPASS study of southern ultradiffuse galaxies and low surface brightness galaxies}

\author[Zhou et al.]{
Yun-Fan Zhou,$^{1,2}$
Chandreyee Sengupta$^{1}$\thanks{Corresponding author: sengupta.chandreyee@gmail.com},
Yogesh Chandola,$^{1}$
O. Ivy Wong,$^{3,4}$
Tom C. Scott,$^{5}$
\newauthor Yin-Zhe Ma$^{6,7,1}$\thanks{Corresponding author: ma@ukzn.ac.za},
 and Hao Chen $^{8,9,1}$
\\
$^{1}$Purple Mountain Observatory (CAS), No. 10 Yuanhua Road, Qixia District, Nanjing 210034, China\\
$^{2}$School of Astronomy and Space Science, University of Science and Technology of China, Hefei, Anhui 230026, China \\
$^{3}$CSIRO Space \& Astronomy, PO Box 1130, Bentley, WA 6102, Australia\\
$^{4}$ICRAR-M468, University of Western Australia, Crawley, WA 6009, Australia\\
$^{5}$ Instituto de Astrof\'{i}sica e Ci\^{e}ncia do Espa\c{c}o (IA), Rua das Estrelas, 4150-762 Porto, Portugal\\
$^{6}$ School of Chemistry and Physics, University of KwaZulu-Natal, Westville Campus, Private Bag X54001, Durban, 4000, South Africa \\
$^{7}$ NAOC-UKZN Computational Astrophysics Centre (NUCAC), University of KwaZulu-Natal, Durban, 4000, South Africa \\
$^{8}$ Research Center for Intelligent Computing Platforms, Zhejiang Laboratory, Hangzhou 311100, China \\
$^{9}$ Department of Astronomy, University of Cape Town, Private Bag X3, 7701 Rondebosch, South Africa
}

\date{Accepted XXX. Received YYY; in original form ZZZ}

\pubyear{2015}

\begin{document}
\label{firstpage}
\pagerange{\pageref{firstpage}--\pageref{lastpage}}
\maketitle

\begin{abstract}

We present results from an H\,{\sc i} counterpart search using the HI Parkes All Sky Survey (HIPASS) for a sample of low surface brightness  galaxies (LSBGs) and ultradiffuse galaxies (UDGs) identified from the  Dark Energy Survey (DES). We aimed to establish the redshifts of the DES LSBGs to determine the UDG fraction and understand their properties. Out of 409 galaxies investigated,  \textcolor {black}{ none were unambiguously detected in \hi\ . Our study was significantly hampered  by the high spectral {\rm rms} of HIPASS and thus in this paper we do not make any strong conclusive claims but discuss the main trends and possible scenarios our results reflect.} The overwhelming number of non-detections suggest that: (A) Either all the LSBGs in the groups, blue or red, have undergone \textcolor{black}{ environment aided pre-processing} and are H\,{\sc i} deficient or \textcolor{black}{the} majority of them are distant galaxies, beyond the HIPASS detection threshold.  (B) The sample investigated is \textcolor{black}{ most likely} dominated by galaxies with \hi\ masses typical of \textcolor{black}{ dwarf galaxies}. Had there been Milky Way (MW) size (\re\ ) galaxies in our sample, with proportionate \hi\ content, they would have been detected, \textcolor{black}{ even with the limitations imposed by the HIPASS spectral quality. This leads us to infer that if some of the} LSBGs have MW size optical diameters, their H\,{\sc i} content is \textcolor{black}{ possibly} in the dwarf range. \textcolor{black}{ More sensitive observations using the SKA precursors in future may resolve these questions.}


\end{abstract}

\begin{keywords}
galaxies: groups: general -- galaxies: evolution -- galaxies: dwarf -- radio lines: galaxies 
\end{keywords}



\section{Introduction}

A relatively unexplored area of extragalactic astronomy is the study of low mass (dwarf) and low surface brightness  galaxies (LSBG).  Understanding the fainter end of the galaxy mass spectrum holds the key to  questions related to galaxy formation, evolution, mass budgets in these structures and thus improving cosmology models. Since their reporting in 2015, a class of fainter LSBGs, called the ultradiffuse galaxies (UDGs; \cite{vdokkum15}) have become a topic of interest to the astronomy community.  To qualify as an UDG, a galaxy has to meet two criteria: they must have a \textcolor{black}{ central surface brightness ($\mu_{\rm g}$) of $\ge$ 24 mag arcsec$^{-2}$ and an effective radius\footnote{The effective radius of a galaxy is the radius at which half of the total light  is emitted} (\re) $\ge$ 1.5 \citep{vdokkum15}.}  While faint, LSBGs are not a recent discovery \citep{impey88, dalc97, conselice18}, the 1000+ UDGs found \textcolor{black}{ projected} around the Coma cluster \citep{koda15} indicated for the first time their relative ubiquity in a dense environments \citep{vaderBurg17}. This fact suggested that UDG studies had the potential to add new insights to knowledge of galaxy and structure formation. Despite the relatively large number of reported UDGs, little is known about their properties and  formation. Various secular and environmentally driven formation scenarios
have been proposed but detailed observations are needed to \textcolor{black}{ determine which ones are valid}.

  UDGs, and LSBGs in general, \textcolor{black}{ are optically faint galaxies with mostly low star formation rates \citep{wyder09}}. As a result, establishing their optical/UV and infrared (IR) properties is observationally expensive. They are typically metal poor, limiting the practicality of molecular gas observations. However, outside cluster cores, UDGs and LSBGs are usually \hi\  rich, making \hi\ line observations a high priority tool to study these galaxies. Despite this,
very few UDG \hi\  studies exist in the literature mainly  \textcolor{black}{ because} the field is new. Single dish targeted \hi\ UDG surveys yielding statistically significant results are so far limited to only a handful of studies \citep[i.e.][]{Leisman17, karunakaran20}. A few more \hi\  studies of UDGs are focused on  \hi\  in isolated UDGs \citep{papastergis17}, \hi\ rich field UDGs \citep{Leisman17} , and UDGs in groups \citep{spekkens18, poulain22}. There are even fewer resolved \hi\ studies of UDGs \citep{sengupta19,mancera19,scott21, gault21, mancera21}. More extensive \hi\ studies of these galaxies is thus timely and relevant as their abundance in different environments has important implications \textcolor{black}{ for} our knowledge of galaxy and large-scale structure formation.

 Using optical imaging from the Dark Energy Survey (DES; \citealt{abbott18}), \cite{tanoglidis21}  reported a large number ($\sim$23790) LSBGs in an area $\sim$ 5000 deg$^{2}$ mainly from the southern hemisphere sky with a fraction of them being UDG candidates. The \cite{tanoglidis21} LSBG catalogue was based on imaging data and thus lacked the essential redshift information necessary to determine the UDG fraction in the catalogue. Unlike the northern  hemisphere where a number of  \hi\ surveys have been carried out, principally with the Arecibo 305m telescope, the \hi\ Parkes All Sky Survey (HIPASS) single dish survey is the only extensive southern \hi\ survey available. Thus HIPASS provides an excellent opportunity to search for \hi\ counterparts to LSBG/UDGs in the \cite{tanoglidis21} catalogue and determine their redshifts.  In this paper we present the results of our search on a subset of southern hemisphere \cite{tanoglidis21} catalogue LSBGs using \hi\ spectra extracted  from HIPASS  data cubes  \citep{barnes01, meyer04, zwaan04, wong06}. We aim to understand what fraction of our sample had detectable \hi\ and their \hi\ properties, and most importantly the fraction of the \hi\ detected LSBGs that qualify as UDGs.

\section{Sample and Methodology}
\subsection{Sample Selection}
\label{sec:selection}
 Our sample was selected from the southern LSBGs in the \citep{tanoglidis21} LSBG catalogue compiled from  Dark Energy Survey (DES) optical imaging. According to the authors' definition, galaxies qualified as LSBGs if they had  g--band effective radii $\ge$ 2.5$^{\prime\prime}$ and a mean surface brightness (in g band) $\ge$24.2 mag arcsec$^{-2}$. While Tanoglidis' LSBGs were found to be distributed all across the southern sky, they also showed projected clustering  around prominent known galaxy groups and clusters. About 80 such \textcolor{black}{ concentrations} were reported in \cite{tanoglidis21}. On the assumption that the  clustering of the LSBGs around known galaxy groups is also true in velocity space, and not just in projection, we selected the LSBGs associated with groups and clusters. Assuming a large dwarf population  dominated this catalogue, we selected primarily nearby groups. We expect that selecting the groups/clusters would provide approximate constraint on the distance to our targets.  While a fraction of LSBGs projected around the groups and clusters could be foreground or background galaxies, choosing nearby groups and clusters increases the probability that the targeted galaxies would be at a similar redshift. As the \hi\ detection threshold increases with  redshift this approach tends to maximise the probability of detecting \hi\ in the LSBGs while minimising the search distance. \textcolor{black}{ A fraction of the reported \cite{tanoglidis21} LSBGs in nearby groups and clusters \textcolor{black}{ are} also UDG candidates.} In defining their UDG sample, those authors followed the standard definition of an UDG, i.e.,  g -- band \re\ $\ge$ 1.5 kpc and the central surface brightness $\mu_{\rm g}$ $\ge$24.0 mag arcsec$^{-2}$ \citep{vdokkum15}. \cite{tanoglidis21} used the distances to the groups or clusters with which they were presumed to be associated to, to estimate the  \re\ \textcolor{black}{ of} the UDG candidates. Detection of  an \hi\ counterpart to these optical  candidates  would thus allow us to determine whether these are truly UDGs.

\begin{figure}
\begin{center}
\includegraphics[width=\columnwidth]{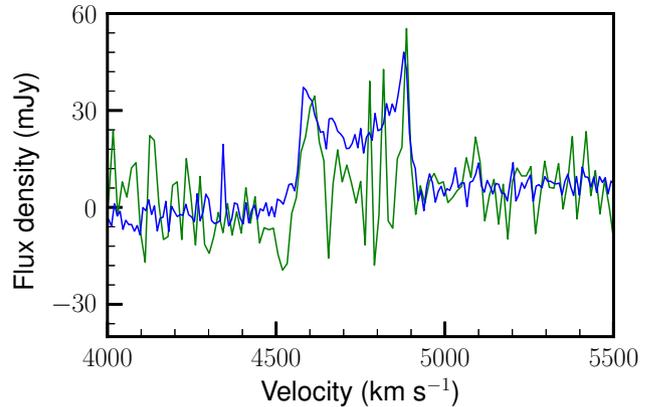}

	\caption{\hi spectra of NGC\,7398, measured from HIPASS (green) and the Arecibo 305m telescope (blue)~\citep{springob05}.}
\label{fig1}
\end{center}
\end{figure}

 We used the HIPASS spectra extracted from the online data release  (https://www.atnf.csiro.au/research /multibeam/release/) to search for \hi\ counterparts in \textcolor{black}{ a} sample 409 of \cite{tanoglidis21} LSBG candidates  with the aim of estimating the \hi\ content of the LSBGs.  \textcolor{black}{ The same exercise was repeated using spectra extracted directly from the HIPASS cubes as a cross check.} Given the HIPASS spectral {\rm rms} $\sim$13 mJy beam$^{-1}$, velocity resolution of 18 \km\ \citep{meyer04} and assuming the \hi\ emission  appears over at least three consecutive channels, a galaxy with an \hi\ mass $\sim 1.9\times 10^{8}\,{\rm M}_{\odot}$, at a distance of 20 Mpc, should be detected at 3$\sigma$ significance with HIPASS. However, had we restricted our sample to distances $\le$ 20 Mpc, our sample size would have been very small. Therefore, we increased our distance limit, being aware that with increasing distance, possibility of detecting galaxies with dwarf \hi\ masses significantly reduces. However, not all LSBGs are  dwarf galaxies and several LSBGs are known to be \hi\  rich and relatively optically extended galaxies \citep{sprayberry95, deblok96, Impey96} and we therefore extended our search to groups with luminosity distances $\le$ 70 Mpc. \textcolor{black}{ At 70 Mpc, a galaxy with an \hi\ mass of $\sim 2.4\times 10^{9}{\rm M}_{\odot}$, still in the dwarf galaxy \hi\ mass  range,  would be  detected at  3$\sigma$ level in a HIPASS spectrum}. Thus, even at 70 Mpc, a few  LSBGs could potentially be detected and thus we included all \cite{tanoglidis21} LSBG candidates in clusters/ groups and overdensities with distances $\le$ 70 Mpc in our sample of 409 LSBGs. Using the archival HIPASS  data, we searched for \hi\ along the line sight for the 409 LSBGs associated with 18 groups and overdensities (15 known groups and 3 central galaxies) with luminosity distances $\le$ 70 Mpc. Table \ref{table1} shows these group names, coordinates, redshift, luminosity distance \textcolor{black}{ as well as the} number of \textcolor{black}{ associated} LSBGs and UDGs (in brackets). The redshifts to these groups/galaxy clusters are taken from \cite{tanoglidis21}. 
 

\begin{table*}
\begin{minipage}{150mm}
	\caption{Groups searched for H{\sc i}. }
	\label{table1}
	\begin{center}
	\begin{tabular}{|l|l|l|l|l|l|l|}
		
		\hline
		(1)&(2)&(3)&(4)&(5)&(6)& (7)\\
		Sl. no.\footnote{Serial number.} &Group/cluster name &R.A.\footnote{All group co-ordinates are from SIMBAD, except RXC J0152.9-1345 and RXC J0340.1-1835 which are from \cite{Piffaretti2011}.}&
		Dec.&
		Redshift\footnote{Redshift of the groups/clusters from \cite{tanoglidis21}.} &
		Lum. dist.\footnote{Luminosity distance of the groups/clusters from \cite{tanoglidis21}}&
		No. LSBGs (UDGs)\footnote{Number of LSBGs (UDGs) in each group/cluster from \cite{tanoglidis21}.}\\
		&&[h\,m\,s]&[d\,m\,s]&&[Mpc]&\\
		\hline
		
		1&Abell S373 (Fornax) & 03:38:30.0 & $-$35:27:18.0 & 0.0046 & 19.0& 59 (3) \\
		2&NGC 1401 & 03:39:21.9 & $-$22:43:29.0 &\textcolor{black}{0.0050}  & 20.3& 26 (1) \\
		3&RXC J0152.9-1345 & 01:52:59.0 &  $-$13:45:12.0& 0.0058 &21.9& 13(0) \\
		4&RXC J0340.1-1835 & 03:40:11.4 & $-$18:35:15.0& 0.0057 & 23.4& 45(1) \\
		5&NGC 1316 &03:22:41.8&$-$37:12:29.5 &\textcolor{black}{0.0059}  & 24.4& 17(1) \\
		6&Abell 3820 &21:52:32.0 &$-$48:23:54.0 & 0.0064 & 25.6& 14(0) \\
		7&NGC 7041 &21:16:32.4&$-$48:21:48.8 & \textcolor{black}{0.0065} & 26.0& 14(1) \\
		8&Abell S989 &22:04:25.0&$-$50:04:24.0 & \textcolor{black}{0.0098} & 40.3& 25(3) \\
		9&NGC 1162 &02:58:56.0&$-$12:23:54.8 & \textcolor{black}{0.0131}  & 55.3& 12(2) \\
		10&NGC 145 &00:31:45.7&$-$05:09:09.6 & \textcolor{black}{0.0138}  & 56.0&10 (0) \\
		11&NGC 829 &02:08:42.2&$-$07:47:26.9 & \textcolor{black}{0.0135}  & 56.1& 17 (5) \\
		12&NGC 1200 &03:03:54.5&$-$11:59:30.7 & \textcolor{black}{0.0135}  & 57.0& 30 (10)\\
		13&Abell 2964 &02:01:06.4&$-$25:04:31.7 & 0.0144  & 60.3& 18 (5)\\
		14&NGC 1521 &04:08:18.9&$-$21:03:07.3 & \textcolor{black}{0.0142}  & 61.4& 14 (4)\\
		15&NGC 1208 &03:06:11.9&$-$09:32:29.4& \textcolor{black}{0.0145}  & 61.6& 18 (5)\\
		16&NGC 199 &00:39:33.2 &+03:08:18.8& 0.0154 & 62.8& 39 (12) \\
		17&NGC 7396 &22:52:22.6&+01:05:33.3& \textcolor{black}{0.0166} & 68.0& 18 (7)\\
		18&Abell S924 &21:07:53.0 &$-$47:10:54.0& \textcolor{black}{0.0162}  & 68.9& 20 (8)\\
		\hline
\end{tabular}

\end{center}
\end{minipage}
\end{table*}

\subsection{Search for \hi\ counterparts and comparison with spectra from the HIPASS cubes}

Lines of sight spectra were extracted from the HIPASS online archive  for each of the 409 LSBGs in our sample in an attempt to detect \hi\ in them. Caveats to this process need to be discussed. The FWHM of the HIPASS beam is large ($\sim$15$^{\prime}$) and in most cases the galaxy coordinates, although within the FWHM of HIPASS beam, differed significantly from the HIPASS beam pointing centre. Additionally, while the canonical {\rm rms} for HIPASS is 13 mJy beam$^{-1}$, depending on sky position it varies from 13 -- 20 mJy beam$^{-1}$ \citep{zwaan04}. These {\rm rms} variations are often convolved with baseline ripples. This fact can add to the difficulty in detecting  galaxies with low \hi\ mass.  The pointing offset and presence of other large group galaxies in the same redshift range within the HIPASS FWHM leads to the risk that the \hi\ signal from our intended target is confused with \hi\ emission from other galaxies within or slightly beyond the HIPASS beam. To minimise this risk for targets associated with groups we restricted our search to only nearby groups ($D \le 70\,{\rm Mpc}$) , while acknowledging that we may have missed several \hi\ counterparts due to this restriction. Figure \ref{fig1} demonstrates the effect of the high spectral {\rm rms} and baseline issues mentioned above. While NGC\,7398 ($D = 67.4\,{\rm Mpc}$), the galaxy in the figure,  is not in our sample, but belongs to one of the groups we are investigating.  It is a large \hi\ rich spiral in contrast to the dwarf dominated LSBG population of our sample. Thus the figure indicates that there is a low probability of detecting our targets with HIPASS unless they are \hi\ rich.

The HIPASS cubes cover a $\sim 8^{\circ}\times 8^{\circ}$ sky area with each pixel covering an area of $\sim 4^{\prime}\times 4^{\prime}$ and the HIPASS FWHM beam is $\sim$15$^{\prime}$ \citep{meyer04}. The spectra available from the website\footnote{\url{https://www.atnf.csiro.au/research /multibeam/release/}} are extracted using a single pixel box at the location of the source, where \textcolor{black}{ the} pixel size is 8$^{\prime}$ $\times$ 8$^{\prime}$. 
\textcolor{black}{ While extracting spectra directly from the HIPASS cubes, we used a $3\,{\rm pixels}\times 3\,{\rm pixels}$ box (with pixel sizes of 4$^{\prime}$), closer to the HIPASS FWHM, for each source. 
We compared the entire set of HIPASS spectra available from HIPASS website to the spectra extracted directly from the cubes. We found no significant difference, however for our analysis we used the spectra \textcolor{black}{ from the $3\,{\rm pixels}\times 3\,{\rm pixels}$} boxes, extracted from the HIPASS cubes.}



\begin{figure}
    \centering 
    \includegraphics[width=0.45\textwidth]{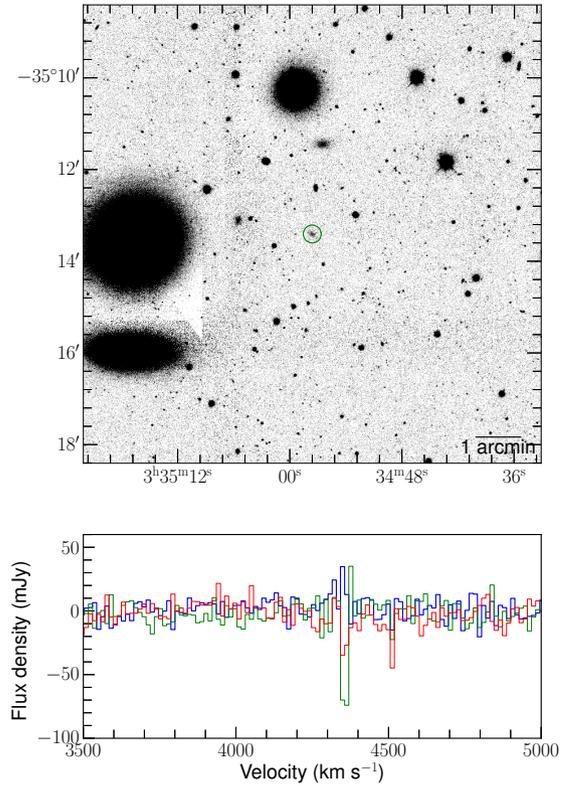}
	\caption{Fornax-C1 (RA 03:34:57.6 Dec -35:13:24.5). {\it Top}--DES image with green circle indicating the galaxy. {\it Bottom}--HIPASS spectrum of the target galaxy in blue. \textcolor{black}{ Spectra in green and red are extracted from regions  1\degree\ away from the galaxy.  The test indicates the peak in this spectrum is RFI. }  }
    \label{fornaxgc1}
\end{figure}


\begin{figure}
    \centering 
    \includegraphics[width=0.45\textwidth]{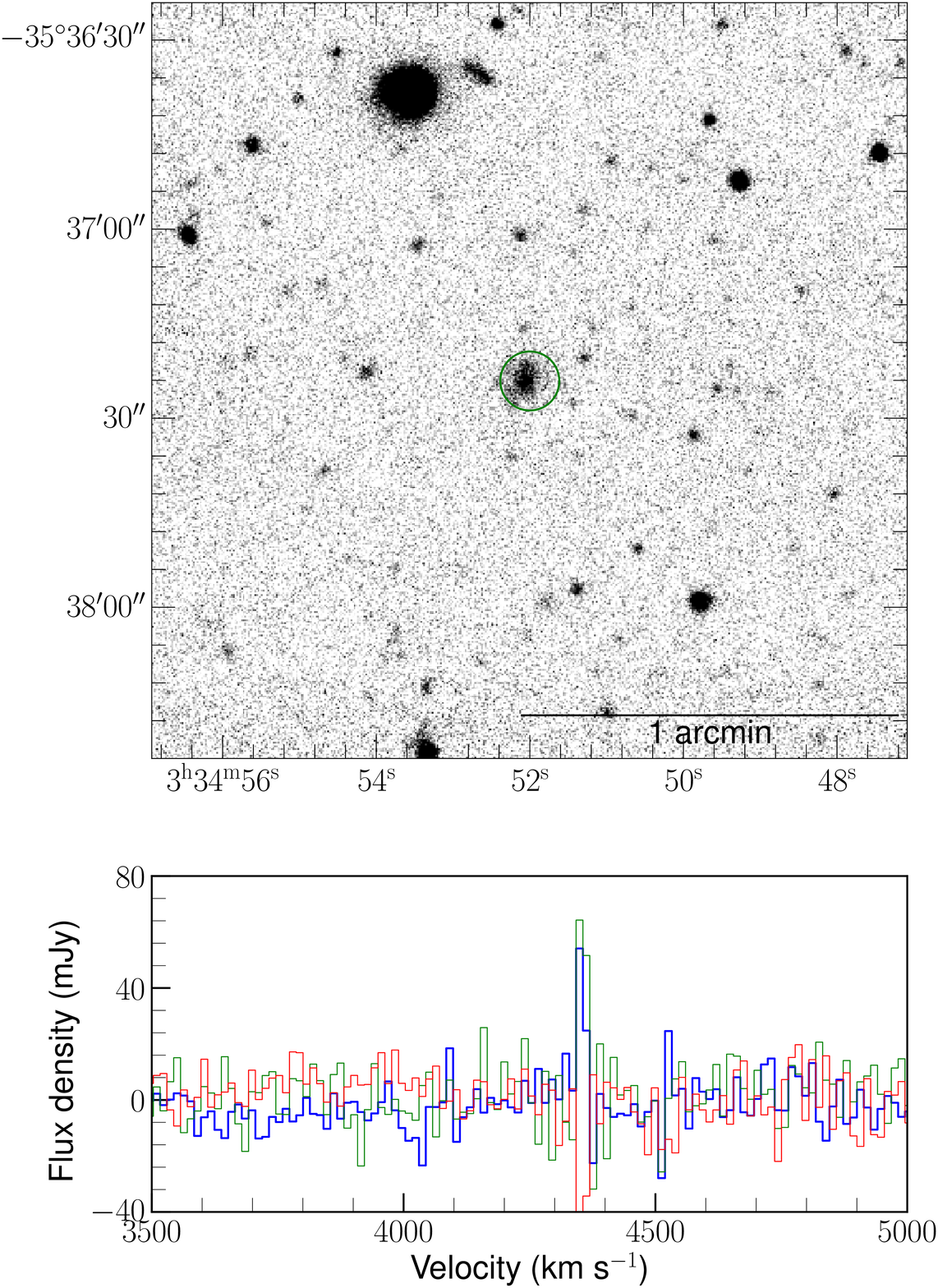}
	\caption{Fornax-C2 (RA 03:34:52.0 Dec -35:37:24.1). {\it Top}--DES image with green circle indicating the galaxy. {\it Bottom}--HIPASS spectrum of the target galaxy. \textcolor{black}{ Spectra in green and red are extracted from regions 1\degree\ away from the galaxy.  The test indicates the peak in this spectrum is RFI. } }
    \label{fornaxgc2}
\end{figure}
\begin{figure}
    \centering
    \includegraphics[width=0.45\textwidth]{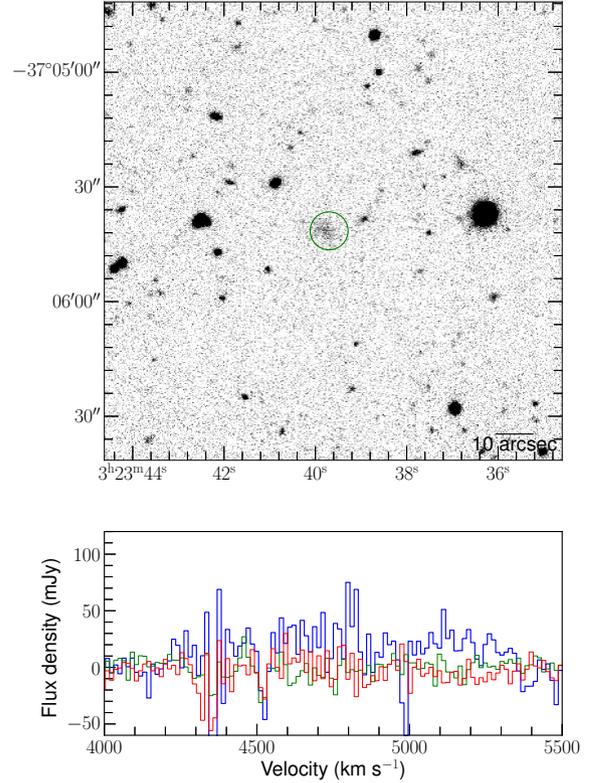}
	\caption{NGC1316-C1 (RA 03:23:39.7 Dec -37:05:41.5). {\it Top}--DES image with green circle indicating the galaxy. {\it Bottom}--HIPASS spectrum of the target galaxy. \textcolor{black}{ Spectra in green and red are extracted from regions 1\degree\ away from the galaxy. The spectrum is barely a 2 sigma signal and cannot be unambiguously claimed as a detection. }}

    \label{ngc1316gc3}
\end{figure}


\begin{figure}
    \centering
     \includegraphics[width=0.45\textwidth]{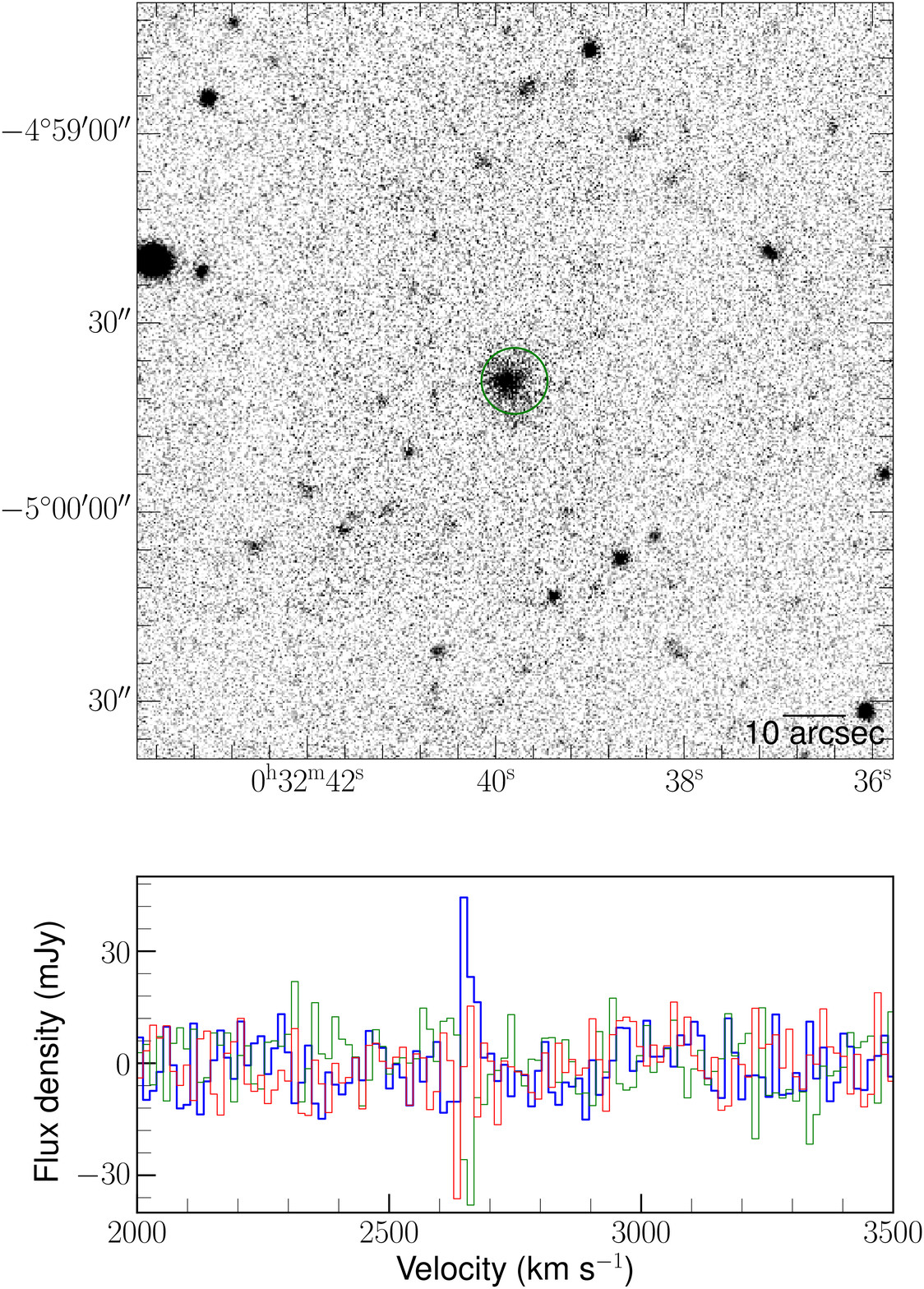}
	\caption{NGC145-C1 (RA 00 32 39.8 Dec -04 59 39.2). {\it Top}--DES image with green circle indicating the galaxy. {\it Bottom}--HIPASS spectrum of the target galaxy.  \textcolor{black}{ Spectra in green and red are extracted from regions 1\degree\ away from the galaxy.  The test indicates the peak in this spectrum is RFI. }}
    \label{ngc145gc1}
\end{figure}


\section{Results}
\textcolor{black}{ Our search for \hi\ in HIPASS cubes for the target galaxies along 409 lines of sight, associated with 18 groups/clusters, yielded no clear detection. There were four tentative detections, two associated with the Fornax cluster and one each with NGC 1316 and NGC 145 groups (see Figures \ref{fornaxgc1}, \ref{fornaxgc2}, \ref{ngc1316gc3} and \ref{ngc145gc1}). The rest were all clear non-detections. All four tentative detections had the following common features. They were all narrow line features, similar to 2 --4 channels and appeared at velocities similar to 4500 \km\ and 2600 \km.  The W$_{20}$ of our tentative detections ranged from 30-50 \km.  Narrow line signals at these frequencies can potentially be \textcolor{black}{ radio frequency interference} {(RFI)}. The HIPASS Data Release Help Page offers information on the frequencies where known RFI signals can be seen. According to this page, the prime interfering line is the 11th harmonic of the 128 MHz sampler clock at 1408 MHz (cz=2640 \km). The page further states, that while this is a narrow line, Doppler corrections may broaden this line by up to 30 \km. Additionally, other residual narrow-band signals may be present in the HIPASS cubes, notably near 1400 MHz, or 4400 \km.  Since some of these RFI signatures match with our tentative detections, we carried out the prescribed RFI checking method suggested in the Data Release Help Page\textcolor{black}{, i.e., by extracting several spectra from along a 1\degree\ radius from the candidate source position.   Of  our tentative sources  Fornax-C1, Fornax-C2 and NGC145-C1,  had narrow  signals from their  1\degree\ radii tests at the same velocity confirming the tentative detections were in fact RFI.}  Sometimes spectrometer saturation may cause a sign bit inversion. This could be a possible reason for seeing negative amplitudes at RFI frequencies. For NGC 1316-C1 we see no such feature. But NGC1316-C1 was the weakest of the four tentative signals and barely a two sigma emission. Thus we conclude we have complete non-detection of \hi\ signals in this search for \hi\ counterparts using the HIPASS data.  We note that a few groups in our sample overlap with the ALFALFA\footnote{\url{http://egg.astro.cornell.edu/alfalfa/data/}} survey areas. Two of our groups (NGC\,199 and NGC\,7369) overlap with the ALFALFA sky coverage, but the LSBGs in those groups were  \hi\ non-detections in both HIPASS and ALFALFA.}

\textcolor{black}{ Assuming the \cite{tanoglidis21} LSBGs to be group members, we next performed a spectral stacking experiment. Due to lack of redshifts for the LSBGs, we assumed all the LSBGs had velocities similar to the nearest group in projection. \textcolor{black}{ Additionally, we only stacked the spectra from the blue galaxies, because these are expected to be \hi\ rich.} The caveat here being that the \textcolor{black}{ groups can have \hi\ velocity dispersions of up to 200 \km\ with group members having a range of radial velocities, whereas the LSBGs are assumed to be at the group systemic velocity. Thus stacking in this case is likely to miss  a major fraction of the galaxies.} However given that these are nearby groups where individual galaxies \textcolor{black}{ with \hi\ masses $\geq$ 10$^{8}$ M$\odot$ should} be detected, the stacked signal would at least detect the LSBGs close to the systemic velocity of the group. Thus the systemic velocity of the host group was considered the zero velocity for all the blue LSBG spectra and a $\pm~$1000 \km\ range about the zero velocity was extracted for stacking them. However, we did not detect any signal in the stacked spectra.}

\textcolor{black}{The \cite{tanoglidis21} LSBG catalogue is based on DES DR1 from the first three years of data from the DES. Their paper contains a link (https://desdr-server.ncsa.illinois.edu/despublic/other\_files/y3-lsbg/) to their LSBG catalogues. We used the original version of the catalogue for our analysis. \textcolor{black}{ But we note that the above website also contains a second version of the catalogue, possibly a recent update on their original version. Comparing the two catalogue versions for our sample showed that galaxies from six groups in our sample were reclassified as field LSBGs rather than group members in version two of the catalogue. The differences between the two versions of the Tanoglidis LSBG catalogues add additional uncertainties to group memberships. But, whether we include or exclude \textcolor{black}{ these six groups, our complete \hi\ non--detection result} remains unchanged \textcolor{black}{ as  do the conclusions.}}}

\section{Discussion}
\textcolor{black}{ Analysis of imaging data from the Dark Energy Survey (DES) provided a large sample of new LSBGs and UDGs mainly in the southern hemisphere \citep{tanoglidis21}. They report a 2D clustering for the red LSBGs where the galaxies appear preferentially near to known groups and clusters. The authors report $\sim$80 such groupings. For a subset of that sample, 18 groups in total, we used the only available large-scale single dish \hi\ survey in the southern hemisphere, HIPASS, to search for \hi\ counterparts. In absence of spectroscopic redshifts, projected proximity to a group or cluster provided the initial distance constraint for our sample.} According to \cite{tanoglidis21}, the majority of the LSBGs associated with over densities are redder than  g -- i$\ge$ 0.60 and the redder LSBGs are more strongly clustered than the bluer ones. This situation introduces a bias in our sample as the bluer galaxies are more likely to be \hi\ detected than the red ones \citep{Leisman17,spekkens18,sengupta19}. \textcolor{black}{However, as a first step, we chose to probe the groups because this provides a better redshift constraint on the sample. \textcolor{black}{ Though rare, it is not impossible for redder LSBGs or dwarfs to contain} substantial \hi\ \citep{Leisman17, papastergis17, karunakaran20, poulain22} and thus we did expect \hi\ detections in at least a fraction of them. In addition, choosing groups does not imply that our sample is completely devoid of blue galaxies. While the dominant population in our 409 LSBG sample have a red colour,  108 are blue galaxies (g--i $<$ 0.6) . }

\textcolor{black}{ Our study \textcolor{black}{ resulted in \hi\ non--detection for all of the}  409 lines of sight in 18 groups. For the HIPASS data, a galaxy's \hi\ mass upper limits ranges from $\sim 1.9\times 10^{8}\,{\rm M}_{\odot}$ (for 20 Mpc) to $\sim 2.4\times 10^{9}{\rm M}_{\odot}$ (for 70 Mpc).} Our \textcolor{black}{ 70 Mpc distance cut off was chosen} to ensure we do not miss higher \hi\ mass but more distant LSBGs, if any. While our best candidates are  projected close to the nearest six groups in our sample (Table \ref{table1}), we extend our distance limit to 70 Mpc. Although, the recently reported UDGs \citep{sengupta19, scott21} are predominantly  dwarf mass galaxies, several LSBGs have been reported to be \hi\ rich with moderate to large size stellar disks \citep{bothun90, sprayberry93}. \textcolor{black}{ So if such galaxies with proportionally large \hi\ masses are present in the \cite{tanoglidis21} LSBG sample, extending the distance limit to 70 Mpc would help us detect them in those more distant groups. Here we discuss a few factors that could explain the \hi\ non-detections in our study.}

According to \cite{tanoglidis21}, of the 409 target LSBGs in our sample, 108 have blue DES color (g -- i $\le$ 0.6) and the majority, 301, are red (g -- i $\ge$ 0.6). While red galaxies can contain detectable \hi\ mass \citep[e.g.][]{Leisman17, papastergis17, karunakaran20, poulain22} at least in the nearby groups, the chances of \hi\ detection in them are lower than bluer galaxies \citep{bouchard05, grossi09,karunakaran20}. Additionally, if these galaxies are genuinely group members, the chance of them being \hi\ deficient is high. \hi\ deficiency from galaxy pre-processing in groups is a known phenomenon and LSBGs with nominal stellar disk mass are  more vulnerable to gas stripping physical processes like tidal interactions, harassment and ram pressure stripping than higher mass galaxies \citep{vm01, seng06, kilborn09, odekon16}. \textcolor{black}{These group physical processes could make even the blue fraction of the LSBGs \hi\ deficient. However, this scenario alone appears insufficient to explain the complete non-detection of the 108 blue galaxies in the sample. Even with pre--processing active in groups, at least a small fraction of the blue galaxies should have been detected at HIPASS sensitivity. \hi\ deficient dwarf galaxies have been detected previously with HIPASS data in groups at similar distances \citep{seng06}.}

An alternative explanation for this non-detections could be that LSBGs, while projected close to the groups, are in fact background galaxies which fall below the HIPASS detection threshold.   \textcolor{black}{ HIPASS's} \hi\ sensitivity falls off rapidly with distance and if a large fraction of our sample are dwarfs and/or in the background of their Tanoglidis assigned group, they would not be detected in the HIPASS. \textcolor{black}{ The result from spectral stacking of the blue galaxies supports this hypothesis. If our LSBGs are group members, statistically at least a fraction of them could have had velocities close to the group systemic velocity.  \textcolor{black}{ Since the groups are at various redshifts, the total blue stacked spectrum rms cannot be used to quote upper limits of \hi\ masses for groups at different distances. Thus individual group's stacked spectral rms was used to extract this number. Thus the 3$\sigma$ upper limit to the \hi\ mass for the nearest ($\sim$20 Mpc) and the farthest ($\sim$70 Mpc) groups are $\sim 3.6\times 10^{7}\msolar$ and  7.3$\times 10^{8}\msolar$ respectively.} For individual galaxies, this limit varies from  $\sim$ 1.9$\times 10^{8}\msolar$ to $2.4\times 10^{9}\msolar$, for the nearest and the farthest groups respectively. These are normal \hi\ \textcolor{black}{ masses for} dwarf galaxies and should have been easily detected in HIPASS, either individually or in the stacked spectra. }

While \textcolor{black}{ our study only results in} non-detections, this exercise, carried out with the best available data at our disposal, \textcolor{black}{ provides a statistical trend for} \hi\ in the \cite{tanoglidis21} LSBGs.  In that context, our results reveal two  important trends.  

\textcolor{black}{ Of our sample of 409 targets, 68 are designated as UDG candidates in \cite{tanoglidis21} and the rest as LSBGs. This classification, however assumes that the galaxies are at the same distances as the groups or clusters they are projected near to. 
Our \hi\ results suggest, a  large fraction of our sample galaxies might not in fact be clustered \textcolor{black}{ near to the groups} they are projected close to. This effect is almost certainly impacting the estimate of  \textcolor{black}{ the true number of} UDGs in the \cite{tanoglidis21} LSBG catalogue. Additionally, our work demonstrates the \textcolor{black}{ critical} importance of spectroscopic observations for these galaxies since redshift confirmation is the only way to understand the true fraction of UDGs in this sample. This result together with the low \hi\ detection rates of UDGs in clusters \citep{karunakaran20} challenge our perceived idea of clustering property of UDGs.}  UDGs are optically selected galaxies and thus the UDG literature is dominated by optical imaging studies~\citep{vdokkum15, koda15, yagi16, roman17, shi17}. They were first reported in the Coma cluster and subsequent reports of their discoveries also came mainly from groups and clusters giving the impression of an enhanced population of these galaxies in such overdensities \citep{vaderBurg17}. \cite{tanoglidis21} also reported a similar clustering for red LSBGs and UDGs in  the southern sky. Our overwhelming number of non-detections, \textcolor{black}{ even for typical \textcolor{black}{ \hi\ mass} dwarf LSBGs or UDGs}, raises doubts about the reported clustering properties.  The 108 blue galaxies in our sample of 409 LSBGs have an even higher probability of being non-cluster or non-group members. This is because galaxies in groups will undergo pre-processing causing gas loss and also redder colour. \textcolor{black}{ Deeper spectroscopic, optical or \hi\, } observations are required to confirm or refute the association of UDGs and LSBGs with the groups/ clusters.

 Our project was designed to detect \hi\ rich LSBGs of all sizes, \textcolor{black}{ including distant \hi\ rich dwarfs} out to a distance of about 70 Mpc. \textcolor{black}{ The} lack of even a single clear detection of \textcolor{black}{ a  LSBG or UDG with the \hi\ mass of the Milky way (MW)  suggested our sample only contains dwarf \hi\ mass  galaxies}. Among the reported UDGs in the recent years, a substantial fraction have  \re\ $\ge$ 3.7 kpc (similar to or larger than that of the MW) \citep{2019ApJS..240....1Z}. The stellar masses of these galaxies may be equivalent to small dwarfs, but their \re\  mimics much larger galaxies. While these UDGs are considerably more extended than dwarf galaxies,  it is not yet clear if the \hi\ line widths, \hi\ masses and the dark matter content are consistent with the dwarf or  more massive galaxies. Recently \cite{gault21} imaged \hi\ in about ten UDGs  and found the  \hi\ mass  and the \hi\ disk diameter to follow the correlation  in \cite{wang16}, however the \hi\ mass range covered in this work is less than $2\times 10^{9}\msolar$,  in the range of dwarf galaxies.  A scaling relation between the UDG \re\ and the DM halo mass was proposed by \cite{Zaritsky17}  and is consistent with a globular cluster count study of six Coma UDGs with \re\ $\ge$ 3 kpc by \cite{saif22}. However the \re\ - DM halo mass relation is yet to be confirmed with DM halo mass estimates based on \hi\ rotation curves. Moreover, if this relation is established for cluster UDGs it is not clear if this would also hold for gas rich field UDGs where the formation mechanism may also be different. 
 
 The lack of \textcolor{black}{spectroscopically confirmed distances}  for our sample makes it impossible to ascertain how many of our target 409 LSBGs have an \re\ $\ge$ 3.7 kpc.   The \textcolor{black}{LSBGs in our sample, with the } largest angular  \re\  are in the range of 14 to 21 $^{\prime\prime}$ \citep{tanoglidis21}.  \textcolor{black}{In the absence of redshift measurements these larger angular \re\ LSBGs  could be at any redshift along the line of sight.} \textcolor{black}{If these larger angular \re\ galaxies, or a fraction of them,}  are at a distance of 70 Mpc then their \re\  would \textcolor{black}{be}  4 -- 7 kpc\textcolor{black}{, i.e. larger than the MW}. \textcolor{black}{ LSBs or more specifically UDGs} with  \re\ \textcolor{black}{ larger than MW} are not unusual and have been detected in \hi\ in \cite{Leisman17}.  Non-detection of even a single extended galaxy (\re\ $\ge$ 3.7 kpc) in our study thus suggests two possibile scenarios: (A) the sample is \textcolor{black}{consists entirely of } LSBGs with \hi\ masses in the range dwarf galaxies and \textcolor{black}{is} devoid of any higher \textcolor{black}{ \re\ } galaxies; (B) \textcolor{black}{ If \textcolor{black}{LSBGs with \re\ $\ge$ the MW  are present in the sample, their non-detection in \hi, suggests that they } have dwarf like \hi\ content and perhaps even dwarf like dark matter content.}

 {Scenario (B) is consistent with recent results from \hi\ studies of faint LSBGs and UDGs. \textcolor{black}{For example} \textcolor{black}{\cite{gault21} studied  a sample of UDGs with \re\ ranging from 1.9 to 6.3 kpc. Irrespective of \re\,  the detected \hi\ mass \textcolor{black}{was} $\le 2\times 10^{9}\msolar$, \textcolor{black}{the \hi\ mass} typically found in dwarf galaxies. \textcolor{black}{The lack \hi\ detections in our study  is consistent with the low \hi\ detection rates in other} studies of UDGs and LSBGs. A recent \hi\ study} of moderately extended (\re $\ge$ 2.5 kpc at the distance of Coma) UDGs from the SMUDGES survey \citep{karunakaran20} resulted in a low detection rate for UDGs.} In that study about 70 UDG candidates were observed using the Green Bank Telescope (GBT) and about 9 UDGs were detected in \hi.  The region surveyed was around the Coma cluster, however none of the \hi\ detected UDGs was cluster members. All of them belong to the low density environment in the foreground or background of the Coma cluster which probably resulted in a better detection rate as opposed to a search inside a group or a cluster, where higher \hi\ deficiencies are expected. Additionally the \hi\ masses of the detected galaxies were $\le 1.7\times 10^{9}\msolar$ irrespective of the \re\, \textcolor{black}{ which again seems to reinforce our findings of Scenario (B) above}.   Compared our 409 targets,  \cite{gault21} and \cite{karunakaran20} had smaller sample sizes, however both of those studies show similar trend to our results with respect to the absence of \hi\ rich and large \re\ UDGs. \textcolor{black}{ While the sample is insufficient to make any strong claims,  Scenario (B) combined with other studies in the literature showing irrespective of \re\, the \hi\ masses of UDGs are typical of dwarf galaxies \citep{gault21,karunakaran20},} most likely suggest that a scaling relation as suggested by \cite{2019ApJS..240....1Z} may not be valid for UDGs. However we clearly need more data and a statistically significant sample to confirm this. \textcolor{black}{The SKA precursors MeerKAT and ASKAP are located in the southern hemisphere. Both telescopes offer higher sensitivity and resolution than HIPASS and therefore could be used in future studies of the LSBGs and UDGs with a higher probability of detecting \hi.}


\section{Conclusions}

Using archival HIPASS  data, we searched for \hi\ counterparts in 409 LSBGs from the~\cite{tanoglidis21} catalogue of southern hemisphere LSBGs.  \textcolor{black}{ We found no convincing \hi\ counterparts for any of the} sample of 409 LSBGs.  

While our study was \textcolor {black}{ significantly hampered } by the high spectral {\rm rms} of HIPASS, the non-detections are not entirely a result of this.  \textcolor {black}{ Our project was designed to detect \hi\ rich LSBGs of all sizes, \textcolor{black}{ including distant \hi\ rich dwarfs} out to a distance of about 70 Mpc. For example, for a distance of 20 Mpc, the HIPASS data would allow us to detect \hi\ mass $\sim 1.9\times 10^{8}\,{\rm M}_{\odot}$ and for 70 Mpc, the farthest group in our sample, the detection limit would be  $\sim 2.4\times 10^{9}{\rm M}_{\odot}$. These numbers represent typical dwarf galaxy, small LSBGs to gas rich small spiral's \hi\ content. Thus a complete non-detection cannot be only due to the limitation of the HIPASS spectral {\rm rms} .}

Our non-detections suggest the following  \textcolor {black}{ likely} scenarios:  \textcolor {black}{ (I)} The majority of LSBGs  are  group members but nearly all of them are \hi\ deficient due to pre--processing in those groups. While many of the red LSBGs could be highly \hi\ deficient and thus below the HIPASS detection limit, this scenario cannot explain the non-detection of all of our sample's 108 blue galaxies.
\textcolor{black}{ (II) Is it possible that our perceived idea of UDG clustering is incorrect. \textcolor{black}{ The majority} of \textcolor{black}{\cite{tanoglidis21}} LSBGs could be distant background galaxies to the groups and thus beyond the detection threshold of the HIPASS. Without more sensitive  spectroscopic measurements this cannot be confirmed. Our study highlights the crucial need for spectroscopy, optical or \hi\ , to estimate the redshifts and to understand \textcolor{black}{ whether}  LSBGs or UDGs are genuine groups members.  \textcolor {black}{ (III)} The sample investigated by us appears to be dominated by galaxies with \hi\ masses in the dwarf range. Had there been  LSBGs or UDGs in our sample with $\ge$ MW \re\ and proportional \hi\ masses, even with the high spectral {\rm rms} of HIPASS, the detection rate would have been higher. We did not even detect any MW \re\ LSBG with an \hi\ mass of the order of a few times 10$^{9} \msolar$ , typically seen in extended UDGs \citep{Leisman17, karunakaran20}. This may imply, LSBGs or UDGs with stellar disks as extended as the MW probably have an  \hi\ content  similar to dwarf galaxies. Clearly more sensitive observations using the SKA precursors in future may answer these questions. }

\section*{Acknowledgements}
\textcolor {black}{ We thank the annonymous referee, whose comments have significantly improved the paper.}  We thank Jayanta Roy and Bhaswati Bhattacharyya of NCRA-TIFR for useful discussions about the paper. The Parkes telescope is part of the Australia Telescope which is funded by the Commonwealth of Australia for operation as a National Facility managed by CSIRO. YC acknowledges the support from the NSFC under grant No. 12050410259, and Center for Astronomical Mega-Science, Chinese Academy of Sciences, for the FAST distinguished young researcher fellowship (19-FAST-02), and MOST for the grant no. QNJ2021061003L. 
TS acknowledges support by Funda\c{c}\~{a}o para a Ci\^{e}ncia e a Tecnologia (FCT) through national funds (UID/FIS/04434/2013), FCT/MCTES through national funds (PIDDAC) by this grant UID/FIS/04434/2019 and by FEDER through COMPETE2020 (POCI--01--0145--FEDER--007672).  TS also acknowledges support from DL 57/2016/CP1364/CT0009. YZM acknowledges the support of National Research Foundation with grant no. 120385 and 120378.  HC is supported by Key Research Project of Zhejiang Lab (No. 2021PE0AC03).

\section*{Data Availability}

This project has used publicly available archived data. 
Koribalski, Baerbel; Staveley-Smith, Lister (2004): The HI Parkes All Sky Survey (HIPASS) image cubes. v1. CSIRO. Data Collection. https://doi.org/10.25919/5c36de6d37141.
The spectra can also be downloaded from https://www.atnf.csiro.au/
research/multibeam/\\release/.



\bibliographystyle{mnras}
\bibliography{cig} 








\bsp	
\label{lastpage}
\end{document}